# Pressure balance at the magnetopause: Experimental studies

A. V. Suvorova[1,2] and A. V. Dmitriev[3,2]

[1]*Center for Space and Remote Sensing Research, National Central University, Jhongli, Taiwan*

[2]*Skobeltsyn Institute of Nuclear Physics Moscow State University, Moscow, Russia*

[3]*Institute of Space Sciences, National Central University, Chung-Li, Taiwan*

**Abstract**

The pressure balance at the magnetopause is formed by magnetic field and plasma in the magnetosheath, on one side, and inside the magnetosphere, on the other side. In the approach of dipole earth's magnetic field configuration and gas-dynamics solar wind flowing around the magnetosphere, the pressure balance predicts that the magnetopause distance $R$ depends on solar wind dynamic pressure $Pd$ as a power low $R \sim Pd^{\alpha}$, where the exponent $\alpha$=-1/6. In the real magnetosphere the magnetic filed is contributed by additional sources: Chapman-Ferraro current system, field-aligned currents, tail current, and storm-time ring current. Net contribution of those sources depends on particular magnetospheric region and varies with solar wind conditions and geomagnetic activity. As a result, the parameters of pressure balance, including power index $\alpha$, depend on both the local position at the magnetopause and geomagnetic activity. In addition, the pressure balance can be affected by a non-linear transfer of the solar wind energy to the magnetosheath, especially for quasi-radial regime of the subsolar bow shock formation proper for the interplanetary magnetic field vector aligned with the solar wind plasma flow.

## A review of previous results

The pressure balance states that the pressure of the flowing around solar wind plasma is balanced at the magnetopause by the pressure inside the magnetosphere [e.g. *Spreiter et al.*, 1966]. As a first approach, the magnetospheric pressure is presented as a pressure of dipole earth's magnetic field $H_d$, which decreases as cube of distance $R$: $H_d \sim R^{-3}$. In a gas-dynamics approach, the solar wind flow is characterized by the dynamic pressure $Pd$. Hence in that approach the pressure balance can be expressed as a proportionality $Pd \sim H_d^2$, i.e. $Pd \sim R^{-6}$. In other words, the magnetopause distance depends on a power function of solar wind dynamic pressure: $R \sim Pd^{\alpha}$, with the exponent $\alpha$=-1/6. *Schield* [1969] proposed more accurate approach and considered a magnetic field generated by the ring current in addition to the geodipole. Later studies consider a net effect of a system of magnetospheric currents, which consists of Chapman-Ferraro current system, tail current, storm-time ring current, and field-aligned currents [e.g. *Stern*, 1995; *Sotirelis and Meng*, 1999]. The magnetic fields generated by those currents have non-dipole configuration. In such approach the power index in the pressure balance is not necessarily equal to -1/6. An additional contribution of magnetospheric plasma thermal pressure to the pressure balance is relatively small and its dynamics is still not well investigated. Contribution of various magnetospheric sources to the pressure balance is an important and intensively studied problem. Experimental studies of the pressure balance were conducted using different satellites with limited data sets at low and high latitudes under quiet and moderately disturbed geomagnetic conditions [*Holzer and Slavin*, 1978; *Bryant and Riggs*, 1989; *Kuznetsov et al.*, 1991; 1992]. Dynamics of low-altitude footprint of cusp, associated with the nose magnetopause distance, was studied using DMSP data [*Kuznetsov et al.*, 1991; 1993]. Most detailed study by *Kuznetsov et al.* found that the power index increases with a conic angle up to 1/7.3~1/10 and in the subsolar region is smaller $\alpha$ = -1/5.3, while others studies got average value close to the theoretical prediction of $\alpha$=-1/6. *Kuznetsov et al.* [1992] estimated the index values depending on location at the magnetopause such as subsolar region, high-latitudes above the cusp, and flanks in the near-earth tail region. At low latitudes, those results have been confirmed on larger statistics [*Kuznetsov et. al.*, 1994, *Kuznetsov and Suvorova*, 1994; 1996], using data set of magnetopause crossings by *Sibeck et al.* [1991]. However, high latitudes suffer from very modest statistics, especially under disturbed conditions accompanied by intensification of the magnetospheric currents. Later, *Boardsen et al.* [2000] filled this gap and presented statistical study of high-latitude magnetopause, antisunward the cusp. Their average estimation of the power index for high latitudes was $\alpha$=-1/7.57 and for nose region was $\alpha$=-1/6.19 .

Comparison of several empirical magnetopause models in the subsolar point [*Kuznetsov and Suvorova* 1996; *Suvorova* 1996; 1997; *Suvorova et al.*, 1999; *Safrankova et al.*, 2002] reveled that different models predict different parameters of the pressure balance such that the model power index $\alpha$ can be larger or smaller than -1/6 by up to ~20%. Modeling the magnetopause with varying power index $\alpha$ [*Suvorova* 1996; *Kuznetsov and Suvorova*, 1998a; *Kuznetsov et al.*, 1998; *Boardsen et al.*, 2000] show that the index is smallest in the subsolar region (~-1/5), increases at the flanks (~-1/6.5) and largest in the tale of magnetopause (see Figure 1). However, most of those studies were based on assumption of axially symmetrical magnetopause shape and did not consider magnetopause asymmetries. The latter can be caused by intensification of magnetospheric currents, which disturb the geodypole magnetic field. A pronounce dawn-dusk magnetopause asymmetry has been found in studies of geosynchronous magnetopause crossings under strongly disturbed solar wind and geomagnetic conditions [*Kuznetsov and Suvorova*, 1997; 1998b; *Dmitriev et al.*, 2004; 2011; *Suvorova et al.*, 2005].

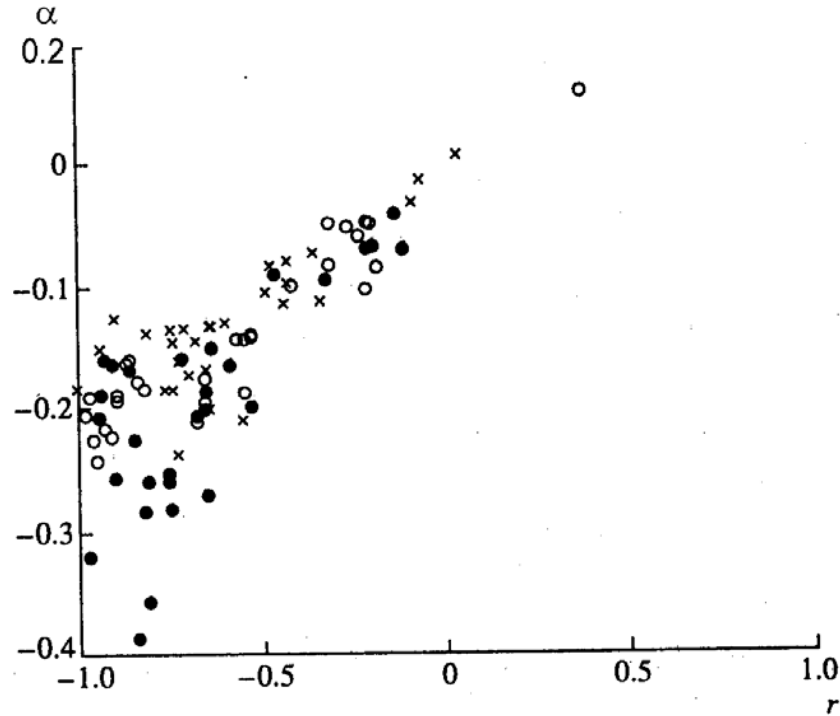

Figure 1. Power index $\alpha$ depending on correlation between the magnetopause distance $R$ and solar wind dynamic pressure for different IMF Bz: white circles (Bz>0), crosses (0>Bz>-3), black circles (Bz<-3) (from *Kuznetsov et al.*, 1998).

It is well known two regimes of the magnetopause formation under northward and southward interplanetary magnetic field (IMF) such that the coefficients in the pressure balance equation are different for the different regimes [e.g. *Aubry et al.*, 1970; *Roelof and Sibeck*, 1993; *Shue et al.*, 1998; *Suvorova et al.*, 1999; 2005]. Modeling the magnetopause by Artificial Neural Network, *Dmitriev and Suvorova* [2000a,b] and *Dmitriev at al.* [1999] show that the power index $\alpha$ decreases for the southward IMF, such that dependence of the magnetopause nose distance on the solar wind pressure becomes weaker. This effect might be a result of intensification of the magnetospheric currents. Recently, *Lin et al.* [2010] have developed a three-dimensional asymmetric magnetopause model on the base of largest historical set of the magnetopause crossings. They have obtained that $\alpha$ = -0.194 ~ -1/5 that is coincides with the findings by *Kuznestov and Suvorova*.

Using MHD approach for modeling the magnetopause under southward IMF, *Wiltberger et al.* [2003] showed that the major driver of magnetopause erosion is the growth of the tail current. *Sotirelis and Meng* [1999] and *Boardsen et al.* [2000] demonstrated the cusp current effect to the subsolar magnetopause. Namely for large tilt angles of the geodypole, the cusp region approaches the equatorial plane and hence influences the subsolar magnetopause shape and distance.

Recent results obtained by Interball, Geotail, CLUSTER and others missions provide new input into progress in studies of the pressure balance. For instance, *Ober et al.* [2006] compared observations with MHD modeling and demonstrated important role of the field-aligned currents intensified during strong southward IMF. They found that under those conditions the dayside magnetopause configuration consists of a reduced dipolar region and sunward extending lobes, such that the magnetopause shape is very blunt. That pattern is formed by decrease of contribution from weakening Chapman-Ferraro current at the magnetopause and the dominating influence of Region 1 field-aligned currents fringing fields. However, the existing results are based on analysis of a few case events that is insufficient for quantitative analysis of the pressure balance at high-latitude and close-tail magnetopause under disturbed conditions.

## New findings

THEMIS mission (2007-2009) gave us tremendous potentialities to continue studying the pressure balance. The subsolar region was considered as a first step of experimental study of pressure balance at the magnetopause. A set of 104 crossings in angular range +/-15° around the Sun-Earth line was matched with upstream solar wind data from ACE and WIND monitors. The pressure balance equation $P \sim R^{-1/\alpha}$ was used to fit the data in 9 subsets corresponding to 3 angular interval of the dipole tilt angle (0-6, 7-15, 15-30°) and 3 intervals of the IMF Bz (Bz <-3; -3<Bz<-1; Bz >-1, nT). It was expected that the fitting in each subset will have high accuracy, i.e. the magnetopause can be well ordered by the GSM location, dipole tilt angle and geomagnetic activity. However, we have found that for nominal condition (Bz >-1 nT) the spread of magnetopause location is unexpectedly large in the range of relatively small values of pressure (<2 nPa) (see Figure 2). That spread could not be reduced by additional binning on tilt, conic angles or Bz value.

A comprehensive analysis of all events from this subset was performed [*Suvorova et al.*, 2010]. A strong relation of the magnetopause position was found with cone angle (defined between the IMF direction and the Sun-Earth line). Under a small cone angles (radial IMF), the magnetosphere experiences a global expansion. This effect causes the largest MP expansion ever documented (about 5 Re in subsolar region). With THEMIS and GEOTAIL data, we prove the global expansion of the magnetopause due to a decrease of magnetosheath pressure in association with radial IMF rather than by solar wind pressure variations. According to THEMIS observation, the magnetosheath pressure applied to the magnetopause was much smaller than the upstream dynamic pressure. The low-pressure magnetosheath (LPM) was contributed mainly by the thermal pressure and was observed in balance with the magnetospheric magnetic field pressure. The problem of the LPM mode is an important and poorly understood issue, which should be a subject of further investigations. The effect of LPM mode results in a weaker dependence of the magnetopause distance from the solar wind pressure, i.e. the power index $\alpha$ should be larger than -1/6 (somewhere ~-1/7).

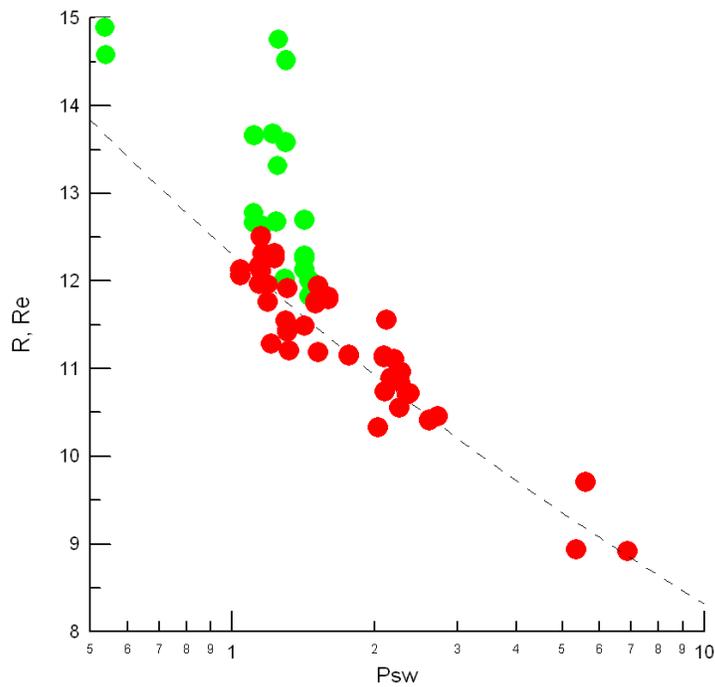

Figure 2. Subsolar magnetopause location versus solar wind dynamic pressure for different IMF orientation: green circles for quasi-radial IMF, red circles - for northward IMF orientation.

Modeling geosynchronous magnetopause crossings occurred mainly during magnetic storms, *Dmitriev et al.* [2011] found that the pressure balance at the subsolar magnetopause is strongly affected by the cross-tail current. A substantial depletion of the dayside geomagnetic field results from the storm-time intensification of the cross-tail current and from sunward motion of its inner edge such that the current approaches to the dayside magnetopause. The cross-tail current magnetic effect decreases with distance slower than that from the dipole ($R^{-3}$). Hence, the net magnetic effect should have a power index larger than -3 (somewhere between -3 and -2), that results in a decrease of the power index $\alpha$ from the theoretical value of -1/6 to empirical $\alpha \sim$ -1/5.